\documentstyle[twocolumn,aps,prb,epsf]{revtex}
\begin{document} 
\twocolumn[\hsize\textwidth\columnwidth\hsize\csname
@twocolumnfalse\endcsname
\widetext
\draft
\title {\bf Observation of Chaotic Dynamics in Dilute 
Sheared Aqueous Solutions of CTAT}
\author{
Ranjini Bandyopadhyay, Geetha Basappa and A. K. Sood
}
\address{ 
Department of Physics, 
Indian Institute of Science,
Bangalore 560 012,
India
}
\maketitle

\begin{abstract}
  
     The nonlinear flow behaviour of a 
viscoelastic gel formed due to entangled, cylindrical micelles in  aqueous solutions of the surfactant CTAT
has been studied.  
On subjecting the system to a step shear rate lying
above a certain value, the  shear and normal stresses show interesting time dependent behaviour.      
The analysis of the measured time series 
shows the existence of a finite correlation dimension and a 
positive Lyapunov exponent, unambiguously implying that the dynamics 
can be described by that of a dynamical system 
with a strange attractor whose dimension increases with the increase in
shear rate.
\vskip0.25cm
PACS numbers: 83.50.By, 83.50.Gd, 83.50.Ws, 82.70.Gg.
\vskip0.25cm

\end{abstract}
]

\narrowtext

Systems of giant wormlike micelles formed in 
certain surfactant solutions are known to show very unusual 
nonlinear rheology. In 
steady shear, the shear stress $\sigma$ saturates to a constant value while the first
normal stress difference increases roughly linearly with shear rate $\dot\gamma$\cite{reh,spe2}.
The 
constitutive model of viscoelastic behaviour of wormlike micellar systems which 
incorporates reptation and reaction dynamics (breakage and recombination 
of micelles), 
predicts a mechanical instability of the {\it shear banding} type  
\cite{spe1} where
bands supporting high shear rates (low viscosity)
coexist with regions of lower shear rates (higher viscosity). 
Flow birefringence \cite{mak} and nuclear magnetic resonance velocity imaging
\cite{mai} have revealed the existence of banded flow 
in the shear stress plateau. An alternative 
explanation for the non-monotonicity of the flow curve has 
also been proposed \cite{ber,por} in
terms of the coexistence of two thermodynamically stable phases - isotropic and 
nematic, present in the sheared solution. Berret \cite{ber} observed
damped, 
periodic oscillations in 
the stress relaxation of CPyCl-NaSal solution at 
a surfactant volume fraction $\phi$ of 12\% on the    
application of controlled shear rates $\dot\gamma$. 
Grand {\it{et al}} \cite{gran} have shown the existence of 
a 
metastable branch in the flow curve of dilute CPyCl-NaSal 
solution supporting stresses higher than that observed in the stress plateau. 
They have explained their results in terms of shear banding. 
Previous observations of shear stress  fluctuations in CTAB-NaSal 
solutions have been explained in terms of 
shear thickening induced by the growth and retraction of shear 
induced structures \cite{pin}. 

The rheology of CTAT (cetyl trimethylammonium tosilate) has been examined    extensively in the linear viscoelastic regime \cite{sol1}. 
Above the Kraft temperature of 23$^\circ$C 
 and at low concentrations (C $<$ 0.04 wt.\%), 
spherical       
micellar solutions are formed which exhibit Newtonian flow behaviour. At        
higher surfactant concentrations (0.04 $<$  C $<$ 0.9 wt.\%), 
cylindrical wormlike micelles are formed which get entangled 
at C $>$ 0.9 wt.\% to form clear 
viscoelastic gels \cite{sol1}.  
The purpose of this Letter is to report interesting time-dependence 
of the shear and normal stresses after subjecting the system to a
step shear rate lying in the plateau region. Our detailed analysis shows unambiguously that the observed 
dynamics can be described as that of a low dimensional, dynamical system
with a strange attractor.

Our experiments were done on dilute aqueous solutions of CTAT, 1.35      
wt.\% at 25$^\circ$C. Following the measurements of the 
elastic modulus G$^\prime(\omega)$ and the viscous modulus  
G$^{\prime\prime}(\omega)$ 
in the linear rheology regime and the flow curve ($\sigma$ vs.$\dot\gamma$) 
in the constant stress mode, 
controlled shear rate experiments were 
performed wherein shear stress and normal stress showed interesting
time-dependence. The time series obtained for both types of stresses
have been analysed 
by two methods to extract certain 
invariant characteristics, metric and dynamical, of the nonlinear dynamics such 
as correlation dimensions, the embedding dimensions and Lyapunov exponents. The
analysis following the algorithm due to Grassberger and Proccacia \cite{gra} 
shows that there is a finite correlation dimension. Another method due to Gao 
and Zheng \cite{gao} yields an estimate of the positive Lyapunov exponent. This 
analysis unambiguously shows that the 
observed time-dependence of the signal is not due to 
stochastic noise, but has its origin in deterministic chaotic dynamics.
We note that Ananthakrishna et al have shown the existence of chaotic dynamics in
the jerky flow  of some metal alloys
undergoing plastic deformation\cite{ana}
which also show a nonmonotonic flow curve.
Recently, after the completion of our present studies \cite{aks},
we became aware of experiments by Soltero {\it{et al}}\cite{sol3} 
on the nonlinear rheology of CTAT at 5 and 10wt\% which show
 the existence of a plateau region at
$\dot\gamma > (2\tau_R)^{-1}$ and 
oscillations in
the stress and birefringence in 5wt\% CTAT 
at $\dot\gamma$=100 s$^{-1}$.

 The CTAT-water samples used in our experiments were prepared by 
adding 
appropriate amounts of CTAT to distilled and deionized water
and were allowed to equilibrate for at least one week. For our 
experiments we used a stress controlled rheometer with temperature control and software for 
shear rate control (Rheolyst AR-1000N, T.A. Instrument, U.K.).
 The rheometer was  also equipped with 
eight strain gauge transducers 
capable 
of measuring the normal force to an accuracy of $10^{-4}$ N.
The measurements were made using a cone-and-plate 
geometry of cone diameter 4 cm and angle 1$^\circ$59$^{\prime\prime}$. 
All the experiments 
reported here have been done on 
fresh samples from the same batch to eliminate sample history effects. The frequency 
response measurements of  
CTAT 1.35wt.\%
by subjecting it to small oscillatory stresses in the linear 
regime reveal a very poor fit to the Maxwell model, as seen in the 
insets (a)
and (b) of Fig. 1, in contrast to 
the behaviour at higher CTAT concentrations  \cite{sol1}.
The reason for the failure 
of the Maxwell model can be that the two relaxation times $\tau_{b}$ and 
$\tau_{rep}$ are not widely different 
($\tau_{b} \sim$ 1.79s as estimated from 
the Cole-Cole plot and $\tau_{rep} \sim$5s obtained from the crossover
frequency $\omega_{CO} =$ 0.3 rad s$^{-1}$ in the G$^{\prime}$ and
G$^{\prime\prime}$ plots, which gives an approximate estimate of the 
relaxation time $\tau_{R}\sim$3s.) We recall that in the Maxwell model,
G(t)=G$_{\circ}$e$^{\frac{-t}{t_{R}}}$, where 
$\tau_{R}=(\tau_{b}\tau_{rep})^{\frac{1}{2}}$, $\tau_{b}$ and 
$\tau_{rep}$ are the breaking and reptation times of the micelles 
with $\tau_{b}<<\tau_{rep}$ \cite{cat}. 

\begin{figure}
\centerline{\epsfxsize = 8cm \epsfbox{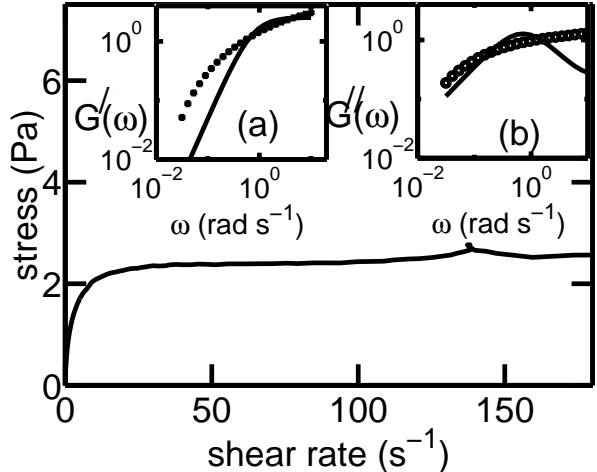}}
\caption{ Metastable flow curve of CTAT 1.35wt.\% at 25$^{\circ}$C,
measured under conditions of controlled stress. Insets (a) and (b) show $G^{\prime}(\omega)$
(shown by squares) and $G^{\prime\prime}(\omega)$ (shown by circles). The solid
lines show the fits to the Maxwell model.}
\end{figure}

   Fig. 1 shows the metastable flow curve measured under conditions of 
controlled stress (time elapsed between acquisition of successive
data points = 2s) for the CTAT solution of concentration 1.35\% by 
weight.  Interestingly, even though CTAT 1.35wt\% is not strictly Maxwellian, its flow
 curve shows a smooth transition to a plateau region above a critical shear rate similar to the ones observed in 
surfactant solutions of CPyCl/NaSal \cite{ber,por,gran} and concentrated
CTAT \cite{sol3}, which show Maxwellian behaviour in G($\omega$). For 10$s^{-1} < \dot\gamma <$ 200$s^{-1}$, 
$\sigma \sim {\dot\gamma}^{\alpha}$ where $\alpha$=0.06 $\pm$ 0.004, implying a very weak concentration dependence\cite{ber2}.
This value of $\alpha$ is very much smaller than that observed in 
the flow curve of the CPyCl/Hexanol/NaCl/water system ($\alpha$=0.3) 
which undergoes an isotropic-nematic transition at $\phi$=31\% \cite{ber2}. 
This observation indicates that the plateau in the flow curve of our system 
is due to shear banding rather than the coexistence of the isotropic and 
nematic phases\cite{ber2}. 
We now discuss our stress relaxation measurements on 
imposing step shear rates of different magnitudes to the samples. 
For low shear rates, $\dot\gamma \leq$ 22 s$^{-1}$, the stress relaxes monotonically to
a steady state value in a few seconds. For higher shear rates,
the stress initially relaxes (t $<$ 100s) and then oscillates in
time, as shown in Fig. 2 (a -e) for a few typical values of applied shear rates.
Fig. 2(f) shows the oscillations in the normal stress measured along
with the shear stress (curve c)
at $\dot\gamma$=100 s$^{-1}$.
On raising the temperature, the oscillations in the stress relaxation are 
found to decrease in amplitude and disappear completely 
at a temperature of 35$^\circ$C (Fig. 2(g)). This may be because of a 
decrease in the 
width of the stress plateau in the flow curve with increasing 
temperatures \cite{por}. The power spectra calculated from the Fourier 
transforms of the stress autocorrelation functions have been 
computed from the time-dependent stress data as shown in Fig. 3.
The initial decay of  
the observed stress (t$<$100s) 
has not been included in the calculations of the power spectrum.

\begin{figure}
\centerline{\epsfxsize = 8cm \epsfbox{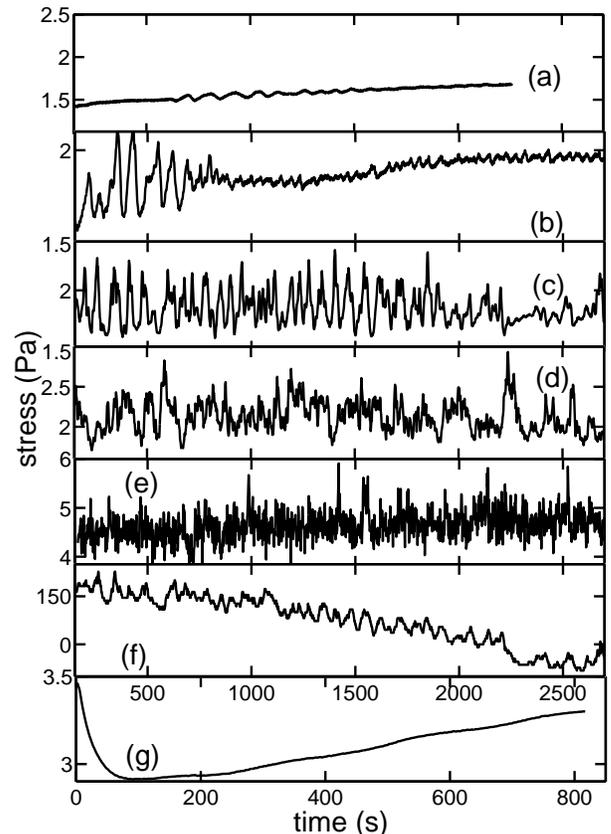}}
\caption{Shear stress relaxation in CTAT 1.35\% on subjecting the sample to
step shear rates of (a) 22.5 s$^{-1}$, (b) 75 s$^{-1}$, (c) 100 s$^{-1}$, (d) 138 s$^{-1}$, 
(e) 175 s$^{-1}$ at 25$^\circ$C. 2(f) shows the time-dependent decay of the normal
stress on application of $\dot\gamma$=100 s$^{-1}$, also at 25$^\circ$C. Figure 2(g)
shows the disappearance of the time-dependent 
oscillations at 35$^\circ$C at $\dot\gamma$=100 s$^{-1}$.}
\end{figure}

 The power spectra 
reveal that for $\dot\gamma$ = 22.5 s$^{-1}$, there is only
one peak at 0.009 s$^{-1}$ corresponding to a time scale of
110 seconds. At $\dot\gamma$ = 75 s$^{-1}$, the power spectrum exhibits 
two dominant peaks at 0.01 and 0.015 s$^{-1}$, corresponding to
time scales of 100 and 67 seconds. At higher shear rates, there are 
many more frequency components. These time scales inferred from 
Fig. 3 may be
understood as due to the realignment of macro-domains which form as a 
result of shear banding. Stick-slip between these domains due to the 
application of shear can result in the observed time-dependent behaviour.

\begin{figure}
\centerline{\epsfxsize = 6cm \epsfbox{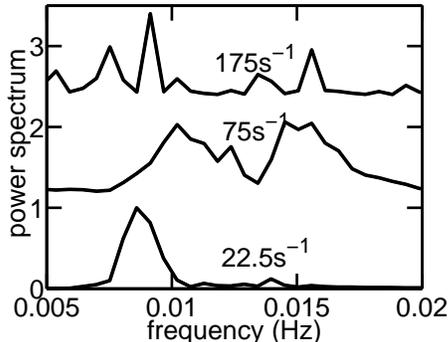}}
\caption{Power spectra of the stress relaxation data at $\dot\gamma$
= 22.5, 75 and 175 s$^{-1}$.}
\end{figure}


   We will now present the analysis of the data to look for any pattern 
expected in a deterministic dynamical system which would distinguish it
from stochastic noise. There is a short time 
predictability in deterministic chaos as compared to no
predictability at all in stochastic noise.
Let $\sigma_j = \sigma (j\Delta t)$ denote the 
time series shown in Fig. 2, consisting of stresses measured at 
regular time intervals $\Delta$t (=1.8s), with j=1 to N (N=1500).
The time series is used to construct an m-dimensional vector 
$\vec X_i=(\sigma_i, \sigma_{i+L}, ..... , \sigma_{i+(m-1)L})$
where m is the embedding dimension and L is the delay time. The 
embedding theorem of Takens  ensures that the dynamics of the
original system is represented by $F: \vec X_i \rightarrow \vec X_{i+1}$, 
provided that the embedding dimension m and the delay time L are 
properly chosen \cite{ott}. A useful way to characterise the dynamical system is by
the correlation dimension $\nu$ \cite{gra} of the (strange) attractor
towards which the phase space trajectories converge in the 
asymptotic limit. The correlation integral C(R) is defined in an 
m-dimensional phase space as 
$C(R)={lim\atop N\rightarrow \infty }{1\over N^{2}}\sum
_{i,j=1}^{N}H({R-|\vec X}_{i}-\vec X_{j}|)$,
where H(x) is the Heaviside functions and $|\vec X_i-\vec X_j|$ is the
distance between the pair of points (i,j) in the m-dimensional
embedding space. The sum in the expression for C(R) 
gives 
the number of point pairs separated by a distance less than R. For small 
R's, C(R) is known to scale as 
$C(R)\sim R^{\nu }$,                           
where  the {\it{correlation dimension}} $\nu$ gives us useful 
information about the local structure of the attractor \cite{gra}. 
The exponent $\nu$ is obtained as a function of log(R) from the plot of
log(C(R)) versus log(R) (Fig. 4). A plateau in the plot of $\nu$ versus
log(R) gives the correct $\nu$ for a chosen embedding dimension
m. If the attractor is unfolded by choosing a large enough
m, then the correlation dimension $\nu$ becomes independent 
of the value of the embedding dimension m. The value of m at 
which this independence sets in is the correct embedding dimension, 
and the corresponding $\nu$ is the correlation dimension.
If an experimental signal 
satisfies $\nu < m$, then the signal is due to deterministic chaos rather than 
random noise\cite{gra}. 

\begin{figure}
\centerline{\epsfxsize = 8cm \epsfbox{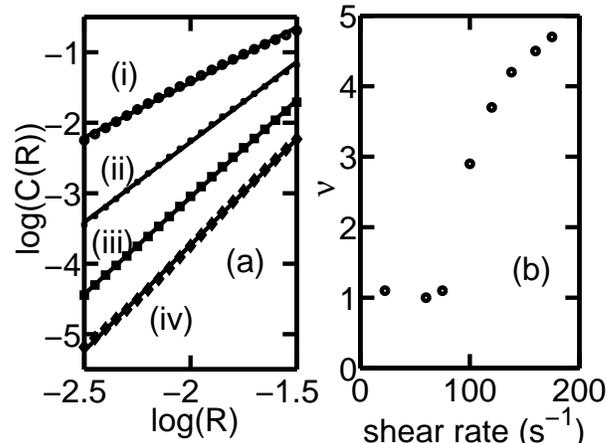}}
\caption{4(a) shows the plot of log(C(R)) vs. log(R) of the stress
trajectories at 
$\dot\gamma$=100 s$^{-1}$ for m=2 to 5 ((i)-(iv)). The slopes of the 
plots give the following values of $\nu$ : (i) $\nu$ =1.6 for m=2, 
(ii) $\nu$ =2.3 for m=3, (iii) $\nu$ =2.8 for m=4 and (iv) $\nu$ =2.8 for
m=5. 4(b) shows the correlation dimensions 
calculated as a function of shear rate $\dot\gamma$.}
\end{figure}

    Fig. 4(a) shows the typical calculations 
of the correlation dimension $\nu$ of   
the attractor to which the stress trajectories asymptotically converge for 
a shear rate of 100 s$^{-1}$.  This gives m$=$4 and $\nu =$2.8. 
Fig. 4(b) reveals a monotonically increasing behaviour of $\nu$ with
$\dot\gamma$.
We see that $\nu$ $\sim$ 1 for
stress relaxation data at $\dot\gamma$ = 22.5 s$^{-1}$ where the power spectrum 
(Fig. 3) shows a single frequency. This is indeed
expected as $\nu$ = 1 for a singly periodic motion
(limit cycle), $\nu$ = 2 for a biperiodic torus
and $\nu$ $>$ 2 for a strange attractor\cite{ott}. 
We see that the correlation dimension $\nu >$ 2
 above a shear rate of 75 s$^{-1}$. The dynamics 
of the stress relaxation thus appears to take 
place on the surfaces of attractors of fractal dimensions that increase with 
the increase in the applied shear rate. The procedure has been repeated for
different data sets acquired at different times and the values of $\nu$ 
calculated in these cases are found to agree to within 15\%. 
We have also calculated the correlation
dimensions of the attractors on which the normal stress orbits lie when 
CTAT 1.35\% was subjected to high shear rates. The normal stress 
is also found to show low-dimensional, deterministic dynamics \cite{ran}

The existence of a positive 
Lyapunov exponent is the most reliable signature 
of chaotic dynamics\cite{ott}. The Lyapunov exponent characterises 
how the distance between two neighbouring points in 
phase space evolves in time. Defining $d_{ij}(k)=||{\vec X_{i+k}-\vec X_{j+k}}||$, the Euclidean 
distance between two vectors constructed using the embedding
theorem by k iterations of the dynamics F, and plotting 
$<ln[d_{ij}(k)/d_{ij}(0)]>$ as a function of k, we calculate
the Lyapunov exponent by using the relation $\lambda = S/\Delta t  ln(2)$, 
where S is the slope of the plot.
It is seen that $\lambda$ becomes positive
at shear rates $>$ 75 s$^{-1}$, indicating the  
onset of chaos at these shear rates (Fig. 5).

\begin{figure}
\centerline{\epsfxsize =8cm \epsfbox{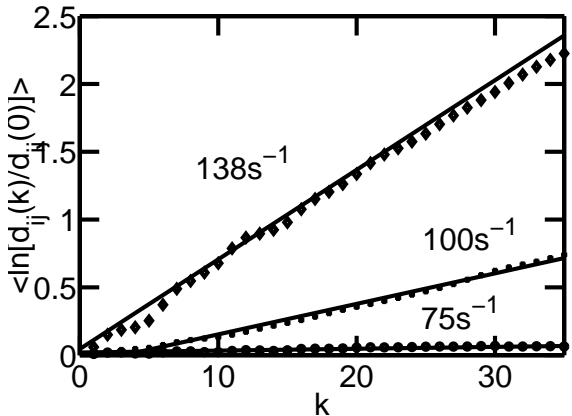}}
\caption{ Calculation of the Lyapunov exponent for $\dot\gamma$=75, 100
and 138 s$^{-1}$. The Lyapunov exponent which may be estimated from the slope 
of the curves is $\sim$ 0 at $\dot\gamma$=75 s$^{-1}$ and becomes positive
at higher shear rates.}
\end{figure}

In order to quantitatively understand our results, we 
need to set up space and time - dependent, nonlinear, 
coupled differential equations in at least four phase space variables. 
These equations will be the equations 
of motion, for example, the 
continuity and momentum balance equations 
together with the constitutive 
relation between viscoelastic stress and shear rate.
The starting point can be the Johnson -Segalman (J-S) 
model \cite{joh} proposed in the context of 'spurt effect'
of polymers which has generically similar 
nonmonotonic behaviour to the reptation - reaction model 
of wormlike micelles. In the J-S model, the viscoelastic nature
of the polymer is accounted for by writing the total stress as the sum of 
a Newtonian part and a deformation history dependent viscoelastic part. 
This model predicts damped oscillations in the
stress in controlled shear rate conditions \cite{ber,ran}.
We believe that 
the coupling of the mean micellar length to the shear 
rate \cite{tur}, the dynamics of the 
mechanical interfaces \cite{yua} and the 
flow-concentration coupling \cite{sch}
should be incorporated in the J-S model.  A model constructed
by incorporating these additional features, which takes into account the 
nonlinear coupling between the relevant dynamical variables 
like shear and normal stresses, shear rate, concentration profiles and 
micellar length distributions
 is most likely to exhibit
the chaotic behaviour that we observe.

In summary, we have proved the existence of 
chaotic dynamics in the rheology of dilute, aqueous solutions of CTAT.
Our analysis unambiguously show that the minimum 
embedding dimension required to describe the shear banding 
instability is four. 
Our experiments and analyses will favour the explanation of the 
nonmonotonic flow behaviour in terms of mechanical instability rather than 
the coexistence of isotropic and nematic phases in the sheared solution. The 
presence of the nematic phase for a very low concentration regime as in our
experiments is highly unlikely. We hope that our experiments will motivate
full theoretical modelling of the shear banding phenomenon.

We thank S. Ramaswamy, V. Kumaran and P. R. Nott for the use of 
the rheometer. RB thanks the CSIR and AKS thanks BRNS, India for financial
support.

\end{document}